\documentclass[prd]{revtex4}
\usepackage{amssymb}
\usepackage{graphicx}
\usepackage{float}

\newcommand{\be}{\begin{equation}}
\newcommand{\ee}{\end{equation}}
\newcommand{\ba}{\begin{eqnarray}}
\newcommand{\ea}{\end{eqnarray}}
\newcommand{\ban}{\begin{eqnarray*}}
\newcommand{\ean}{\end{eqnarray*}}

\begin{document}

\title{Complete plasmon spectrum of two-stream system}

\author{Katarzyna Deja}
\affiliation{National Centre for Nuclear Research, Warsaw, Poland}

\author{Stanis\l aw Mr\' owczy\' nski}
\affiliation{Institute of Physics, Jan Kochanowski University, Kielce, Poland \\
and National Centre for Nuclear Research, Warsaw, Poland}

\date{June 20, 2015}

\begin{abstract}

The complete spectrum of plasmons of the two interpenetrating plasma streams is found in a closed analytic form. The orientation of the wave vector with respect to the stream direction is arbitrary and the plasmas, which are assumed to be collisionless and spatially homogeneous, can be nonrelativistic, relativistic or even ultrarelativistic. Our results apply to the electromagnetic plasma of electrons and passive ions and to the quark-gluon plasma governed by QCD. 

\end{abstract}



\maketitle

\section{Introduction}

A plasma system, which is a complex of particles and fields, supports a wide variety of waves. Using quantum terminology one says that the plasmas reveal rich spectra of collective excitations or modes. The spectra carry information about the thermodynamic and transport properties of equilibrium plasmas and about the temporal evolution of non-equilibrium systems. The waves or modes corresponding to oscillations of charge density or currents are usually of the highest frequency and thus they play a particularly important role in the plasma dynamics. In the case of electromagnetic plasmas we deal with electric charges and currents and the electromagnetic waves are classical representation of {\em quasiphotons}. In the quark-gluon plasma  governed by QCD there are color charges and color currents and the {\em quasigluon} is the analog of the quasiphoton. We call the collective modes related to the oscillations of charges and currents as {\em plasmons}, not limiting this term to the equilibrium plasma. 

Plasmons, which are found as solutions of dispersion equations, have been extensively studied for various plasma configurations over decades and there is a huge literature about the problem, see {\it e.g.} the textbooks \cite{Krall-Trivelpiece-1973,Akhiezer-1975}.  However, there is a rather limited number of exact analytic solutions of dispersion equations and there are only exceptional cases when the complete spectrum of plasmons can be found in a closed analytic form. We present here a general and exact solution of the dispersion equation of the two-stream system for an arbitrary orientation of the wave vector with respect to the stream direction. The plasma system is collisionless and spatially homogeneous, but it can be nonrelativistic, relativistic or even ultrarelativistic. Our results apply to both electromagnetic and quark-gluon plasmas. The complete spectrum, which is given in a closed analytic form, consists of eight modes (four pairs of modes of opposite sign) which are either real or imaginary.  For any orientation of the wave vector, there is one unstable mode. When the wave vector is perpendicular to the stream direction, we have the magnetic filamentation mode which continuously changes into the electrostatic longitudinal mode when the wave vector becomes parallel to the stream. 

Two real modes exhibit the phenomenon of mode coupling,  see {\it e.g.} the handbook \cite{Landau-Lifshitz-1981}, when the wave vector is almost perpendicular to the stream or the stream velocity approaches the speed of light. Then, the two dispersion curves approach each other but they do not cross. The effect is encoded in the solutions of the dispersion equation we found and to reveal it no additional reasoning or approximate methods, which are typically used \cite{Landau-Lifshitz-1981}, are not needed. The whole spectrum is rather rich and nontrivial and it changes qualitatively with the orientation of the wave vector.

The two-stream plasma system, which occurs in various plasma physics experiments, is repeatedly discussed in the literature,  see {\it e.g.} the textbooks \cite{Krall-Trivelpiece-1973,Akhiezer-1975}. In context of the quark-gluon plasma it was first studied in \cite{Pokrovsky:1988bm,Mrowczynski:1988dz}. However, as far as we know, the complete spectrum of plasmons in such a system has not been found in closed analytic form and thus our results are new and original. And since analytic form of the spectrum greatly simplifies derivation of various plasma characteristics, the results are also practically useful.

Presentation of our analysis is organized as follows. In the forthcoming Sec.~\ref{sec-form} we first formulate the problem by writing down the general dispersion equation of plasmons. Then, the momentum distribution of plasma constituents, which controls the dielectric tensor, is briefly discussed and an explicit expression of the tensor, which enters the dispersion equation, is derived. In  Sec.~\ref{sec-decom-ABCD} we introduce the method to solve the dispersion equation which is of the form $\det[\Sigma]=0$. Instead of computing the determinant we rather invert the matrix $\Sigma$ and look for poles of $\Sigma^{-1}$. An analysis of the collective modes starts in  Sec.~\ref{sec-special-cases-k-u} with the discussion of two special cases.  Then, we derive the complete set of exact solutions of general dispersion equations in Sec.~\ref{sec-general-case}. The spectrum of plasmons is found for any orientation of the wave vector. Sec.~\ref{sec-special-case-u2=1} discusses the case, which is interesting both physically and mathematically, when the stream velocity approaches the speed of light. In Sec.~\ref{sec-conclusions} we summarize our study and make some final remarks. 

Our analysis of the two-stream system is methodologically very close to the extensive study of plasmons in the system where the momentum distribution of plasma constituents is obtained from the isotropic one by stretching or squeezing it in one direction \cite{Carrington:2014bla}. Since the two-stream distribution does not belong to this category, the technique to invert the matrix $\Sigma$ needs to be modified. However, there are some repetitions with the article \cite{Carrington:2014bla} which are unavoidable to make the present paper self-contained. We use notation very similar to that in Ref.~\cite{Carrington:2014bla} with the natural units, where $\hbar = c =1$, and the indices $i,j,k = 1, 2, 3$ which label the Cartesian spatial coordinates. Lengths of vectors like ${\bf k}$ or ${\bf u}$ are denoted as $k$ and $u$. 

\section{Formulation of the problem}
\label{sec-form}

The linearized chromodynamic or Maxwell equations of the Fourier transformed (chromo-)electric field can be written in the form
\be
\label{maxwell-1}
\Sigma^{ij}(\omega,{\bf k})E^j(\omega,{\bf k})=0 ,
\ee
where the matrix $\Sigma$ is defined as
\be
\label{matrix-Sigma}
\Sigma^{ij}(\omega,{\bf k}) \equiv
- {\bf k}^2 \delta^{ij} + k^ik^j 
+ \omega^2 \varepsilon^{ij}(\omega,{\bf k}) ,
\ee
$\omega$ is the frequency, ${\bf k}$ denotes the wave vector and $\varepsilon^{ij}(\omega,{\bf k})$ is the (chromo-)dielectric tensor, see {\it e.g.} the review article \cite{Mrowczynski:2007hb}.  External charges and currents are absent in the system under study and color indices, if needed, are suppressed in Eqs.~(\ref{maxwell-1}, \ref{matrix-Sigma}). A solution of Eq.~(\ref{maxwell-1}) exists if
\be
\label{general-dis-eq-det}
{\rm det}[ \Sigma(\omega,{\bf k}) ] = 0 ,
\ee
which is the general dispersion equation. Its solutions $\omega({\bf k})$, which represent plasmons, are, in general, complex but the wave vector ${\bf k}$  is assumed to be real. There are the {\it transverse} plasmons, for which the electric field is transverse to the wave vector ${\bf k}$, and {\it longitudinal} plasmons with the electric field parallel to ${\bf k}$. The transverse modes correspond to oscillations of current, and the longitudinal ones to oscillations of charge density.  A mode is called {\it unstable} if $\Im \omega ({\bf k}) > 0$, because the amplitude $\sim \! e^{\Im\omega ({\bf k}) \, t}$ grows exponentially in time. When $\Im \omega ({\bf k}) \le 0$, the mode is {\it stable} and it is {\it damped} if $\Im \omega ({\bf k}) < 0$, as its amplitude decays exponentially in time. The mode is called  {\it overdamped}, when additionally it is pure imaginary.

To solve the dispersion equation, the dielectric tensor is needed. As discussed in {\it e.g.} \cite{Mrowczynski:2007hb}, the tensor for a locally chargeless anisotropic plasma in the collisionless limit equals 
\be
\label{eij}
\varepsilon^{ij}(\omega,{\bf k}) = \delta^{ij} +
\frac{g^2}{2\omega} \int {d^3p \over (2\pi)^3} \,
\frac{v^i}{\omega - {\bf v}\cdot {\bf k}+i0^+}
\Big(\big(1-\frac{{\bf k}\cdot {\bf v}}{\omega}\big) \delta^{jk}
+ \frac{v^jk^k}{\omega} \Big) \nabla_p^k f({\bf p}) ,
\ee
where ${\bf p}$, $E_{\bf p}$ and ${\bf v} \equiv {\bf p}/E_{\bf p}$ are the momentum, energy and velocity of plasma constituents and  $f({\bf p})$ is their distribution function. In case of electromagnetic plasma the coupling constant $g$ should be replaced by $e$ but the results are then in the so-called Lorentz-Heaviside units not in the Gauss units typically used in the plasma physics. For the quark-gluon plasma governed by QCD with the ${\rm SU}(N_c)$ gauge group and one quark flavor, $f({\bf p})= n({\bf p})+ \bar n({\bf p}) +2N_c n_g({\bf p})$, where $n({\bf p})$, $\bar n({\bf p})$, $n_g({\bf p})$ are the distribution functions of quarks, antiquarks and gluons of a single color component. The chromodielectric tensor does not carry any color indices, as the state corresponding to the momentum distribution $f({\bf p})$ is assumed to be colorless. The $i0^+$ prescription makes the Fourier transformed dielectric tensor $\varepsilon^{ij}(t,{\bf r})$ vanish for $t<0$, which is required by causality. In kinetic theory, the infinitesimal quantity  $i0^+$ can be treated as a remnant of inter-particle collisions. Performing the integration by parts, the dielectric tensor (\ref{eij}) can be rewritten in the form
\be
\label{eij-1}
\varepsilon^{ij}(\omega,{\bf k}) = \delta^{ij} -
\frac{g^2}{2\omega^2} \int {d^3p \over (2\pi)^3} \,
\frac{f({\bf p})}{E_{\bf p}}
\bigg[\delta^{ij} +
\frac{k^i v^j + v^i k^j}
{\omega - {\bf v}\cdot {\bf k}+i0^+}
+ \frac{({\bf k}^2 - \omega^2) v^i v^j}
{(\omega - {\bf v}\cdot {\bf k}+i0^+)^2} \bigg] ,
\ee
which will be more convenient for our purposes than the expression (\ref{eij}). 

The dielectric tensor given by Eq.~(\ref{eij}) or (\ref{eij-1}) is fully determined by the momentum distribution of plasma constituents. The distribution function of the two-stream system is chosen to be
\be
\label{f-2streams}
f({\bf p}) = (2\pi )^3 \rho 
\Big[\delta^{(3)}({\bf p} - {\bf q}) + \delta^{(3)}({\bf p} + {\bf q}) \Big] ,
\ee
where $\rho$ is the effective density of plasma constituents in a single stream. The distribution function (\ref{f-2streams}) can be treated as an idealization of the two-peak distribution where the particles have momenta close to ${\bf q}$ or $-{\bf q}$ but it is not required that the momenta are exactly ${\bf q}$ or $-{\bf q}$. 

The distribution function (\ref{f-2streams}) substituted into Eq.~(\ref{eij-1}) provides the dielectric tensor which in turn gives the matrix (\ref{matrix-Sigma}) of the form
\be
\label{sigma-2-stream}
\Sigma^{ij}(\omega,{\bf k}) \equiv
(\omega^2- {\bf k}^2-\mu^2) \delta^{ij} + k^i k^j
- \frac{\mu^2 ({\bf k} \cdot {\bf u}) }{\omega^2 - ({\bf k} \cdot {\bf u})^2} (k^i u^j+u^i k^j)
- \frac{\mu^2(\omega^2+({\bf k} \cdot {\bf u})^2)({\bf k}^2-\omega^2) }{(\omega^2-({\bf k} \cdot {\bf u})^2)^2}u^iu^j ,
\ee
where $\mu^2 \equiv g^2 \rho/ E_{\bf q}$ is a parameter analogous to the Debye mass squared, and ${\bf u} \equiv {\bf q}/E_{\bf q}$ is the stream velocity. It is smaller than the speed of light, if plasma constituents have non-vanishing mass. If we consider a system of massless constituents, $E_{\bf q} = |{\bf q}|$ and ${\bf u}^2 =1$. However, when the distribution (\ref{f-2streams}) is treated as an approximation of the two-peak structure and the particles have non-zero momenta perpendicular to the stream velocity, $E_{\bf q} < |{\bf q}|$ and ${\bf u}^2 < 1$.  We also note that $\mu$ is the only dimensional parameter which enters the problem. Therefore, all dimensional quantities can be expressed in the units set by the appropriate power of the mass $\mu$. 

Our objective is to find a complete set of solutions of the dispersion equation (\ref{general-dis-eq-det}) with the matrix $\Sigma$ given by Eq.~(\ref{sigma-2-stream}). It is not difficult to compute the determinant and solve the  dispersion equation (\ref{general-dis-eq-det}) for specifically chosen orientations of the wave vector. In this way we will consider two special cases in Sec.~\ref{sec-special-cases-k-u}. In general, however, it appears much easier to invert the matrix $\Sigma$ and look for poles of $\Sigma^{-1}$ than to find zeros of the determinant of $\Sigma$. Therefore, we invert the matrix $\Sigma$ in the subsequent section.

\section{Poles of $\Sigma^{-1}$}
\label{sec-decom-ABCD}

To invert the matrix $\Sigma$, one expresses the matrix together with the inverse matrix $\Sigma^{-1}$ in a given basis. Then, the inverse matrix is found directly from the equation  $\Sigma \Sigma^{-1}=1$. To choose the appropriate basis, one observes that the matrix $\Sigma$ given by Eq.~(\ref{sigma-2-stream}) is symmetric $(\Sigma^{ij} = \Sigma^{ji})$ and that it depends on two vectors ${\bf k}$ and ${\bf u}$. Therefore, the matrix can be completely decomposed in the basis of four projectors built out of the vectors  ${\bf k}$ and ${\bf u}$ and the unit matrix. Following Romatschke and Strickland \cite{Romatschke:2003ms}, we introduce the vector ${\bf u}_T$ which is a component of ${\bf u}$ transverse to ${\bf k}$ that is
\be
\label{nT-def}
u_T^i = \big(\delta^{ij} - \frac{k^i k^j}{{\bf k}^2}\big) \, u^j ,
\ee 
and we define four projectors 
\be
\label{A-B-C-D}
A^{ij} \equiv 
\delta^{ij} - \frac{k^i k^j}{{\bf k}^2},
\;\;\;\;\;\;\;
B^{ij} = 
\frac{k^i k^j}{{\bf k}^2} ,
\;\;\;\;\;\;\;
C^{ij} \equiv
\frac{u_T^i u_T^j}{{\bf u}_T^2},
\;\;\;\;\;\;\;\;\;\;
D^{ij} \equiv 
k^i u_T^j + k^j u_T^i .
\ee
The matrix $\Sigma$ is expressed in this projector basis as
\be
\label{Sigma-A-B-C-D}
\Sigma^{ij} = a\,A^{ij} +b\,B^{ij} +c\,C^{ij} +d\,D^{ij}\, ,
\ee
with the coefficients $a, \; b, \; c, \; d$ provided by the equations
\ba
\label{a-b-c-d}
k^i \Sigma^{ij} k^j = {\bf k}^2 b , \;\;\;\;\;
u_T^i \Sigma^{ij} u_T^j = {\bf u}_T^2 (a + c) , \;\;\;\;\;
u_T^i \Sigma^{ij} k^j = {\bf u}_T^2 {\bf k}^2 d , \;\;\;\;\;
{\rm Tr}\Sigma = 2a + b + c .
\ea
With the matrix $\Sigma$ given by Eq.~(\ref{sigma-2-stream}), the coefficients $a, \; b, \; c, \; d$ equal
\ba
\label{a}
a(\omega,{\bf k}) &=& \omega^2 - \mu^2 -{\bf k}^2 ,
\\[2mm]
\label{b}
b(\omega,{\bf k}) &=& \omega^2 - \mu^2 
- \frac{2\mu^2  ({\bf k}\cdot {\bf u})^2}
{\omega^2 - ({\bf k}\cdot {\bf u})^2}
- \frac{\mu^2 \big(\omega^2 + ({\bf k}\cdot {\bf u})^2\big)
({\bf k}^2 - \omega^2)}
{\big(\omega^2 - ({\bf k}\cdot {\bf u})^2\big)^2}
\frac{({\bf k}\cdot {\bf u})^2}{{\bf k}^2} \;,
\\[2mm]
\label{c}
c(\omega,{\bf k}) &=& 
-\frac{\mu^2(\omega^2+(\mathbf{k}\cdot\mathbf{u})^2)({\bf k}^2-\omega^2)}{(\omega^2-(\mathbf{k}\cdot\mathbf{u})^2)^2}  
\left({\bf u}^2 -\frac{(\mathbf{k}\cdot {\bf u})^2}{{\bf k}^2}\right) ,
\\[2mm]
\label{d}
d(\omega,{\bf k}) &=& 
-  \frac{\mu^2  ({\bf k}\cdot {\bf u})}
{\omega^2 - ({\bf k}\cdot {\bf u})^2} 
- \frac{\mu^2 \big(\omega^2 + ({\bf k}\cdot {\bf u})^2\big)
({\bf k}^2 - \omega^2)({\bf k} \cdot {\bf u})}
{{\bf k}^2 \big(\omega^2 - ({\bf k}\cdot {\bf u})^2\big)^2} .
\ea
When compared to the matrix decomposition applied in the studies \cite{Carrington:2014bla} and \cite{Romatschke:2003ms}, there is one important difference: the vector ${\bf u}$, which is analogous to the vector ${\bf n}$ from \cite{Carrington:2014bla} and \cite{Romatschke:2003ms}, is not of unit length but ${\bf u}^2 \le1$. Therefore, the expression ${\bf u}^2$ shows up in the formula (\ref{c}). 

Expressing the inverse matrix $\Sigma^{-1}$ in the same basis (\ref{A-B-C-D}) and solving the equation $\Sigma \Sigma^{-1}=1$, one finds the inverse matrix in the following form
\ba
\label{dispXX}
(\Sigma^{-1})^{ij} =
\frac{1}{a} \,A^{ij} 
+ \frac{-a(a+c)\,B^{ij} 
+ (- d^2{\bf k}^2{\bf u}_T^2 +bc)\,C^{ij}
+ad \,D^{ij}}
{a(d^2{\bf k}^2{\bf u}_T^2 - b(a+c))} .
\ea
As seen, the poles of the matrix $\Sigma^{-1}(\omega,{\bf k})$ are determined by the equations 
\ba
\label{dis-eq-A}
a(\omega,{\bf k}) &=& 0 ,
\\[2mm]
\label{dis-eq-G}
b(\omega,{\bf k}) \big( a(\omega,{\bf k}) + c(\omega,{\bf k}) \big)-{\bf k}^2 {\bf u}_T^2 d^2(\omega,{\bf k})
 &=& 0 ,
\ea
which are the two dispersion equations equivalent to the general dispersion equation (\ref{general-dis-eq-det}). 

The dispersion equation (\ref{dis-eq-A}) with the coefficient $a(\omega,{\bf k})$ given by the formula (\ref{a}) has the simple solution 
\be
\label{A-mode}
\omega^2_a({\bf k}) = \mu^2 + {\bf k}^2 ,
\ee
which represents transverse plasmon. A real problem is to solve the dispersion equation  (\ref{dis-eq-G}).

\section{Special cases ${\bf k} \perp  {\bf u}$ and ${\bf k} \parallel  {\bf u}$}
\label{sec-special-cases-k-u}

Before considering a general solution of the dispersion equation (\ref{dis-eq-G}), we discus here two special cases ${\bf k} \perp {\bf u}$ and  ${\bf k} \parallel {\bf u}$ which will help us to analyze the general case.

\subsection{${\bf k} \perp  {\bf u}$ }
\label{sec-k-perp-u}

When $\theta$, which is the angle between ${\bf k}$ and ${\bf u}$, equals $90^\circ$ that is ${\bf k}\cdot {\bf u} = 0$, the coefficients (\ref{a}-\ref{d}) simplify to
\ba
\label{a-k-perp-n}
a(\omega,{\bf k}) &=& \omega^2 - \mu^2 - k^2 ,
\\[2mm]
\label{b-k-perp-n}
b(\omega,{\bf k}) &=& \omega^2 - \mu^2 ,
\\[2mm]
\label{c-k-perp-n}
c(\omega,{\bf k}) &=& 
\frac{m^2 (\omega^2 - k^2)u^2} {\omega^2} ,
\\[2mm]
\label{d-k-perp-n}
d(\omega,{\bf k}) &=& 0 .
\ea
The dispersion equation (\ref{dis-eq-G}) reads
\be
(\omega^2 - \mu^2)\Big(\omega^2 - \mu^2 - k^2 
+ \frac{\mu^2 (\omega^2 - k^2)u^2} {\omega^2}\Big) = 0 ,
\ee
and it has three solutions 
\be
\label{solutions-k-perp-n}
\omega_0^2(k) = \mu^2 ,\;\;\;\;\;\;\;
\omega_{\pm}^2(k) = 
\frac{1}{2}\Big( \lambda^2 + k^2 \pm \sqrt{ \big(\lambda^2 + k^2\big)^2+4\mu^2 u^2 k^2}\;\Big) ,
\ee
where $\lambda \equiv \mu \sqrt{1 - u^2}$. So, there are three pairs of modes of opposite sign. 

To clarify a physical character of the solutions (\ref{solutions-k-perp-n}), we explicitly compute the matrix $\Sigma$ assuming that ${\bf u} = (0,0,u)$ and ${\bf k} = (k,0,0)$. Then, one finds
\ba
\label{matrix-Sigma-k-perp-n}
\Sigma(\omega,{\bf k}) = \left[
\begin{array}{ccc}
\omega^2 - \mu^2 & 0 & 0 
\\
0 & \omega^2 - \mu^2 -k^2 & 0
\\
0 & 0 & 
\omega^2 - \mu^2 - k^2 
+ \frac{\mu^2(\omega^2 - k^2)u^2}{\omega^2}
\end{array}
\right]
\ea
and its determinant equals
\be
{\rm det}[\Sigma(\omega,{\bf k}) ] = 
(\omega^2 - \mu^2)(\omega^2 - \mu^2 - k^2)
\Big(\omega^2 - \mu^2 - k^2 
+ \frac{\mu^2(\omega^2 - k^2)u^2}{\omega^2}\Big) .
\ee
The general dispersion equation (\ref{general-dis-eq-det}), that is ${\rm det}\Sigma = 0$, gives, as expected, the solutions (\ref{solutions-k-perp-n}) and additionally (\ref{A-mode}). The structure of the matrix (\ref{matrix-Sigma-k-perp-n}) clearly shows that the mode $\omega_0(k)$ is longitudinal (the electric field is along the wave vector) and the remaining modes $\omega_a(k), \;\omega_\pm(k)$ are transverse (the electric field is perpendicular to the wave vector). The solutions $\omega_a^2(k)$, $\omega_0^2(k)$ and $\omega_+^2(k)$, which are all positive, correspond to stable real modes while the solution $\omega_-^2(k)$, which is negative, represent two imaginary modes - the Weibel or filamentation unstable and overdamped modes. Let us also note that the solutions  $\omega_0^2(k)$ and $\omega_+^2(k)$ cross each other at $k = \mu u/\sqrt{1 + u^2}$.

\subsection{${\bf k} || {\bf u}$ }
\label{sec-k-para-u}

When the decomposition (\ref{Sigma-A-B-C-D}) is used to invert the matrix $\Sigma$, the case  ${\bf k} || {\bf u}$ needs some care as then ${\bf u}_T = 0$. The fact that the vectors ${\bf k}$ and ${\bf u}$ are parallel to each other means that the matrix $\Sigma$ actually depends on one vector only. Indeed, for $\theta = 0$, when $({\bf k}\cdot {\bf u})^2 = k^2 u^2$, the matrix equals
\be
\label{Sigma-k-parallel-u}
\Sigma^{ij}(\omega,{\bf k}) = 
(\omega^2 - \mu^2 - k^2) \delta^{ij} 
+ \bigg( 1 - \frac{2 \mu^2 u^2 } {\omega^2 -  k^2 u^2}
-\frac{\mu^2 u^2 (\omega^2 + k^2 u^2)
( k^2 - \omega^2)}{k^2 (\omega^2 -  k^2 u^2 )^2}  
\bigg) k^ik^j .
\ee
Consequently, one needs only the matrices $A$ and $B$ to fully decompose $\Sigma$ {\it i.e.}
\be
\label{Sigma-A-B}
\Sigma = a\,A + b\,B ,
\ee
where the coefficients $a$ and $b$, which are found from the equations
\ban
k^i \Sigma^{ij} k^j = k^2 b , \;\;\;\;
{\rm Tr}\Sigma = 2a + b ,
\ean
equal
\ba
\label{a-k-parallel-n-reduced}
a(\omega,{\bf k}) &=& \omega^2 - \mu^2 - k^2
\\[2mm]
\label{b-k-parallel-n-reduced}
b(\omega,{\bf k}) &=& \omega^2 - \mu^2  
- \frac{2 \mu^2 k^2 u^2 }{\omega^2 - k^2 u^2}
-\frac{\mu^2 u^2 (\omega^2 + k^2 u^2) (k^2 - \omega^2)}
{(\omega^2 - k^2 u^2 )^2}  .
\ea
The inverse matrix is
\be
\Sigma^{-1} = \frac{1}{a} \,A + \frac{1}{b} \,B .
\ee
We again have the dispersion equation $a(\omega,{\bf k})=0$, which gives the solution  (\ref{A-mode}), and instead of the dispersion equation (\ref{dis-eq-G}) we have $b(\omega,{\bf k}) = 0$ which gives longitudinal modes, as $B$ projects on the direction parallel to ${\bf k}$. 

With the coefficient $b$ given by the formula (\ref{b-k-parallel-n-reduced}), the dispersion equation is 
\be
\label{eq-dis-long}
\omega^2\Big(
\omega^4  - \big(2 k^2 u^2  + \mu^2 (1-u^2) \big) \omega^2 
+ k^4 u^4 - \mu^2 k^2 u^2 (1-u^2)\Big) = 0 ,
\ee
which, except the trivial solution $\omega^2 =0$, has two solutions 
\be
\label{solution-L-parallel-u}
\omega_\pm^2(k) = k^2 u^2 + \frac{\lambda^2}{2} 
\pm \frac{\lambda}{2} \sqrt{8 k^2 u^2 + \lambda^2},
\ee
where, as previously, $\lambda \equiv \mu \sqrt{1 - u^2}$. As seen, $\omega_+^2(k)$ is always positive but $\omega_-^2(k)$ is negative for $ k^2 < \lambda^2/u^2$. Then, we have the instability which is well known in the plasma physics as the two-stream electrostatic instability. When $u^2 \rightarrow 1$, the mode $\omega_-(k)$  becomes stable and $\omega_-^2(k) =\omega_+^2(k) = k^2$.

The solutions (\ref{A-mode}) and (\ref{solution-L-parallel-u}) can be also easily found directly from the matrix $\Sigma$. Choosing ${\bf u} = (0,0,u)$ and ${\bf k} = (0,0,k)$ the matrix equals
\ba
\Sigma = \left[
\begin{array}{ccc}
\omega^2 - \mu^2 - k^2 & 0 & 0 
\\
0 & \omega^2 - \mu^2 - k^2 & 0
\\
0 & 0 & \omega^2 - \mu^2 - \frac{2\mu^2 k^2 u^2}{ \omega^2 -  k^2 u^2}
-  \frac{\mu^2 (\omega^2 +  k^2 u^2) (k^2 - \omega^2)  u^2}{ (\omega^2 -  k^2 u^2)^2}
\end{array}
\right] .
\ea
The equation ${\rm det}\Sigma = 0$ has, as expected, the solutions (\ref{A-mode}) and (\ref{solution-L-parallel-u}) and the former one is doubled.

\section{General case}
\label{sec-general-case}

\begin{figure}[t]
\begin{minipage}{8.5cm}
\center
\includegraphics[width=1.05\textwidth]{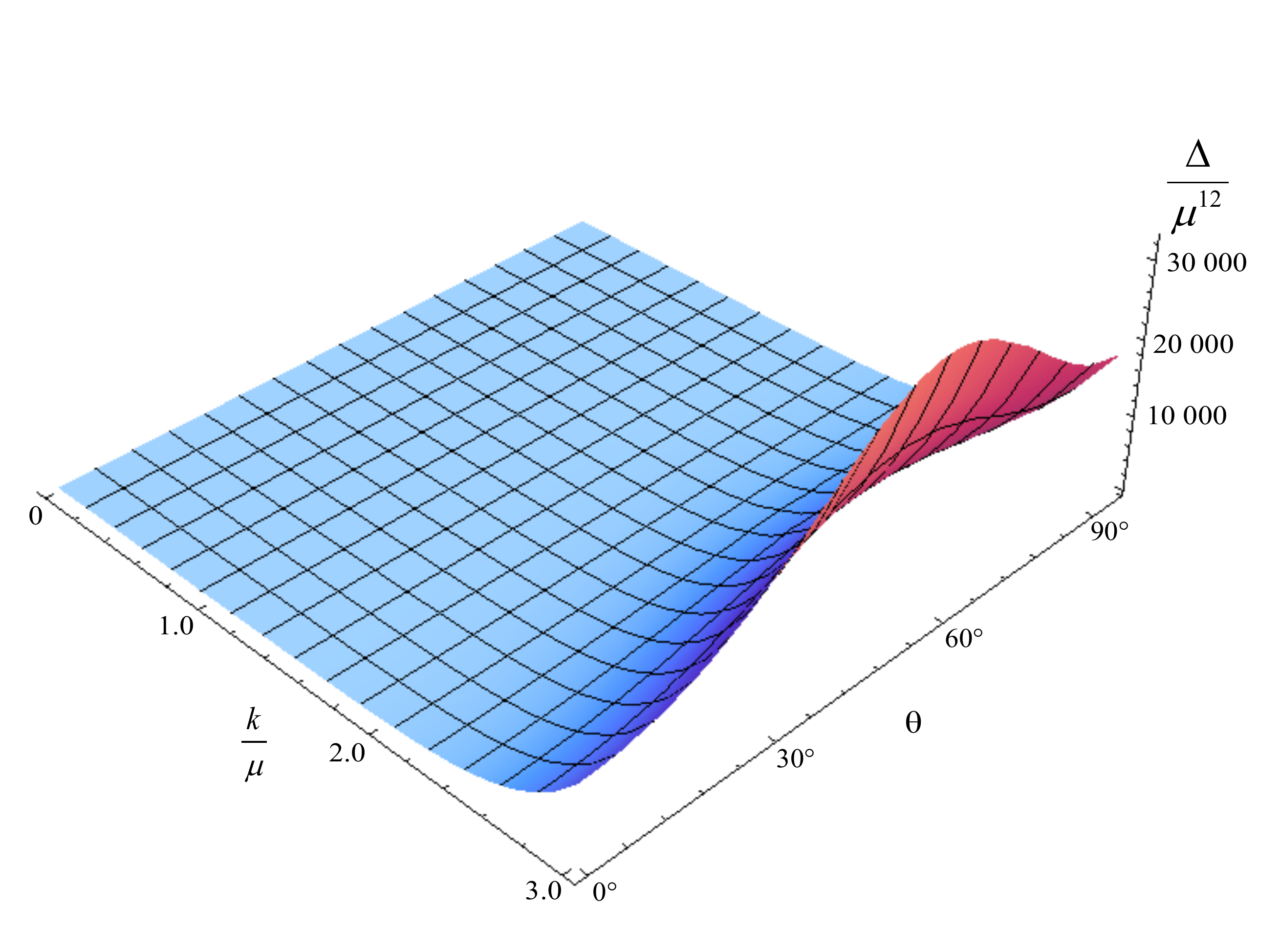}
\end{minipage}
\hspace{1mm}
\begin{minipage}{8.5cm}
\center
\includegraphics[width=1\textwidth]{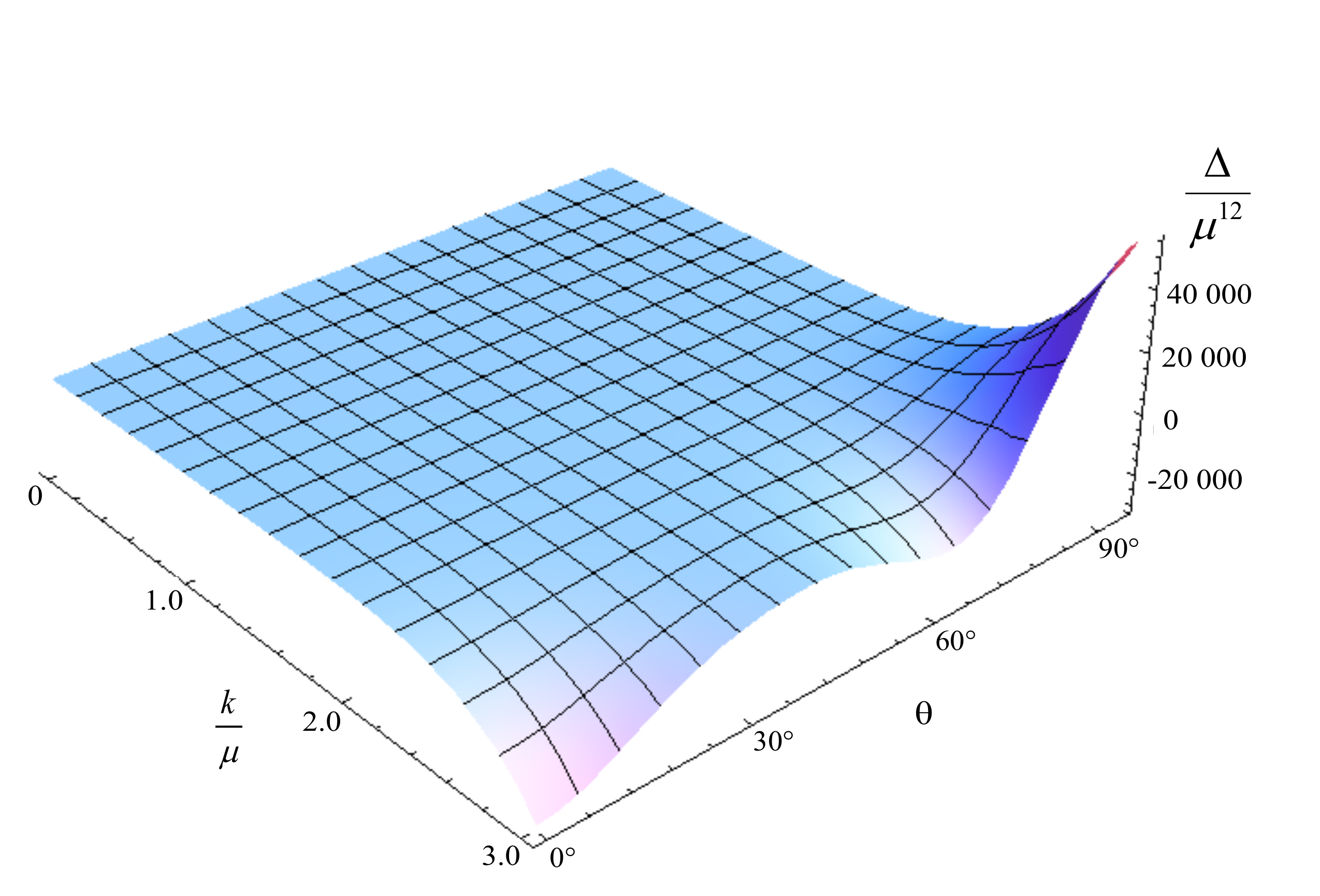}
\end{minipage}
\caption{(Color online) The discriminant $\Delta$ as function of $k$ and $\theta$ for ${\bf u}^2= 1/2$ (left panel) and ${\bf u}^2= 3/2$ (right panel). 
\label{fig-Delta}}
\end{figure}

We discuss here in full generality the dispersion equation (\ref{dis-eq-G}) with the coefficients $a, \; b, \; c, \; d$ given by  Eqs.~(\ref{a}-\ref{d}). A crucial finding is that the ratio $\omega^2/\big(\omega^2 - ({\bf k}\cdot {\bf u})^2\big)^2$ factors out in left-hand-side of the equation. Consequently,  we get the cubic dispersion equation 
\ba
\label{cubic-eq}
a_3 x^3+ a_2 x^2+ a_1 x+a_0 = 0,
\ea
where $x \equiv \omega^2$ and the coefficients $a_0, \, a_1, \, a_2, \, a_3$ are the real numbers equal to
\ba
\label{a_0}
a_0 &=&-\big( \mu^2{\bf u}^2 + ({\bf k}\cdot {\bf u})^2 \big)
\Big(\mu^2({\bf k}\cdot {\bf u})^2+{\bf k}^2 \big( ({\bf k}\cdot {\bf u})^2-\mu^2\big)\Big),
\\[1mm]
\label{a_1}
a_1 &=& ({\bf k}\cdot {\bf u})^2\big(({\bf k}\cdot {\bf u})^2+2{\bf k}^2+\mu^2(1+{\bf u}^2)\big)
+\mu^2(1-{\bf u}^2)({\bf k}^2+\mu^2) ,
\\[1mm]
\label{a_2}
a_2 &=& - {\bf k}^2 - 2({\bf k}\cdot {\bf u})^2 + \mu^2(-2+{\bf u}^2) ,
\\[1mm]
\label{a_3}
a_3 &=& 1.
\ea

As well known, see {\it e.g.} \cite{Bronshtein-Semendyayev-1985}, all three roots of a cubic equation can be found algebraically. Since the coefficients  $a_0, \, a_1, \, a_2, \, a_3$ are real, the character of the roots depends on a value of the discriminant 
\ba
\label{discriminant}
\Delta=18 \, a_0 a_1 a_2 a_3 - 4 \, a_2^3 a_0 + a_1^2 a_2^2 - 4 \, a_3 a_1^3 - 27 \, a_0^2 a_3^2.
\ea
One distinguishes three cases:
\begin{itemize}

\item if $\Delta > 0$, the roots are real and distinct;

\item if $\Delta = 0$, the roots are real and at least two of them coincide;

\item if $\Delta < 0$,  one root is real and the remaining two are complex.

\end{itemize}

One shows that the discriminant (\ref{discriminant}) computed with the coefficients (\ref{a_0}-\ref{a_3}) is nonnegative for any ${\bf k}$, if $0 \le {\bf u}^2 \le 1$ but there is a domain of ${\bf k}$ where $\Delta$ is negative for ${\bf u}^2 > 1$. This is demonstrated in Fig.~\ref{fig-Delta} where the discriminant is plotted as a function of $k$ and $\theta$ for ${\bf u}^2 = 1/2$ (left panel) and ${\bf u}^2 = 3/2$ (right panel). Since the stream velocity is limited by the speed of light, we always have three real solutions of the dispersion equation  (\ref{cubic-eq}). 

\begin{figure}[t]
\begin{minipage}{8.5cm}
\center
\includegraphics[width=\textwidth]{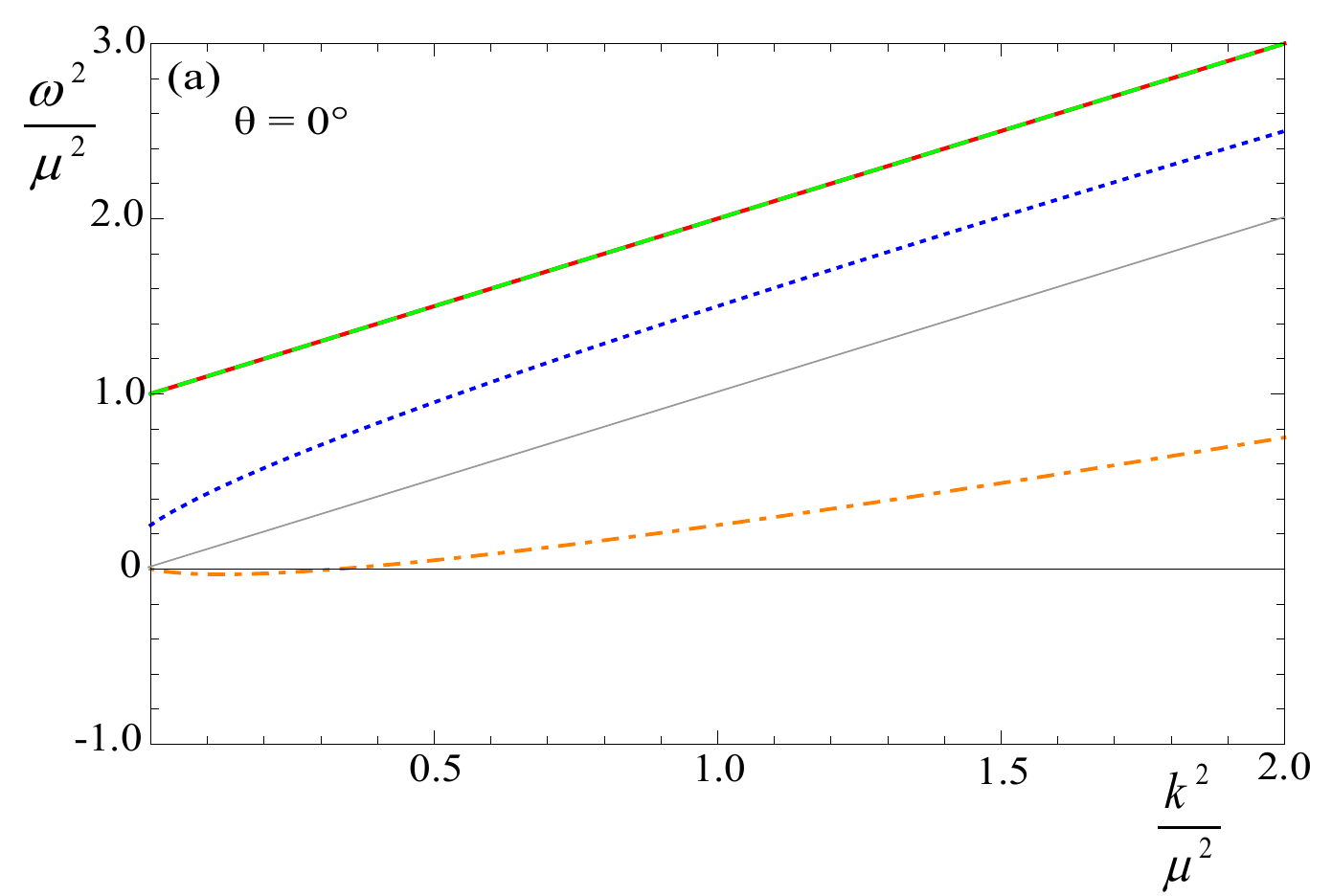}
\end{minipage}
\hspace{1mm}
\begin{minipage}{8.5cm}
\center
\includegraphics[width=\textwidth]{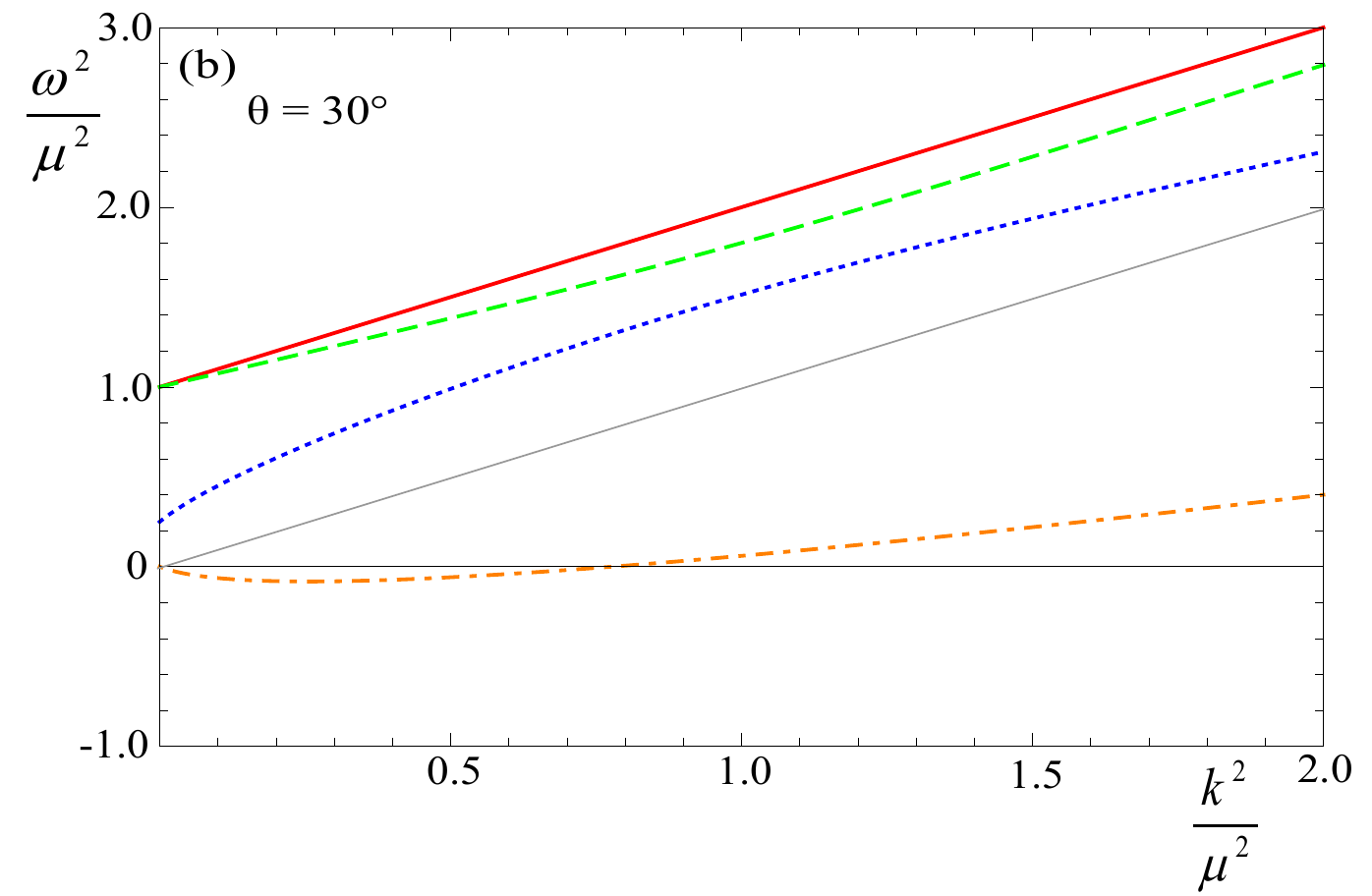}
\end{minipage}
\begin{minipage}{8.5cm}
\center
\includegraphics[width=\textwidth]{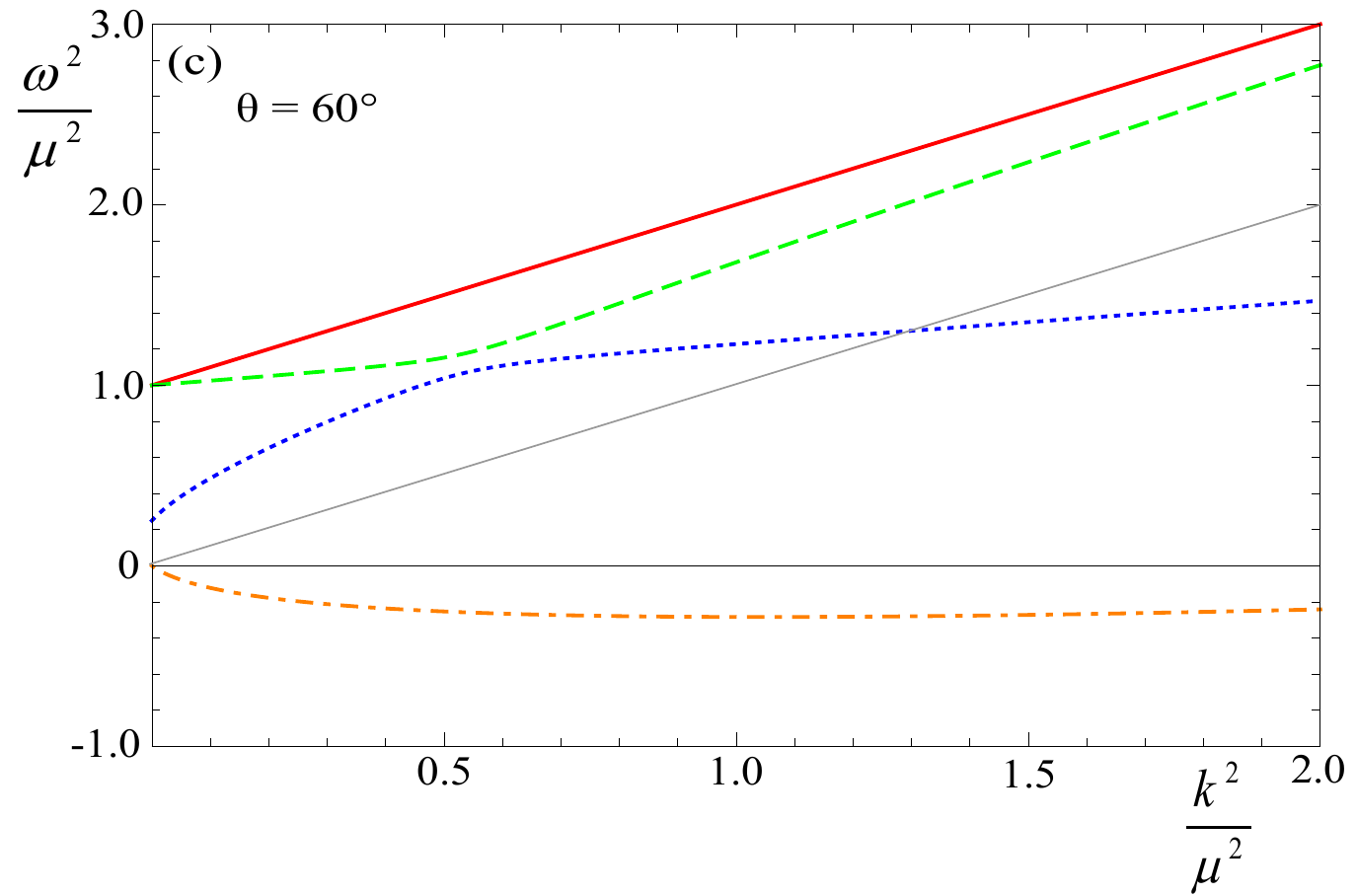}
\end{minipage}
\hspace{1mm}
\begin{minipage}{8.5cm}
\center
\includegraphics[width=\textwidth]{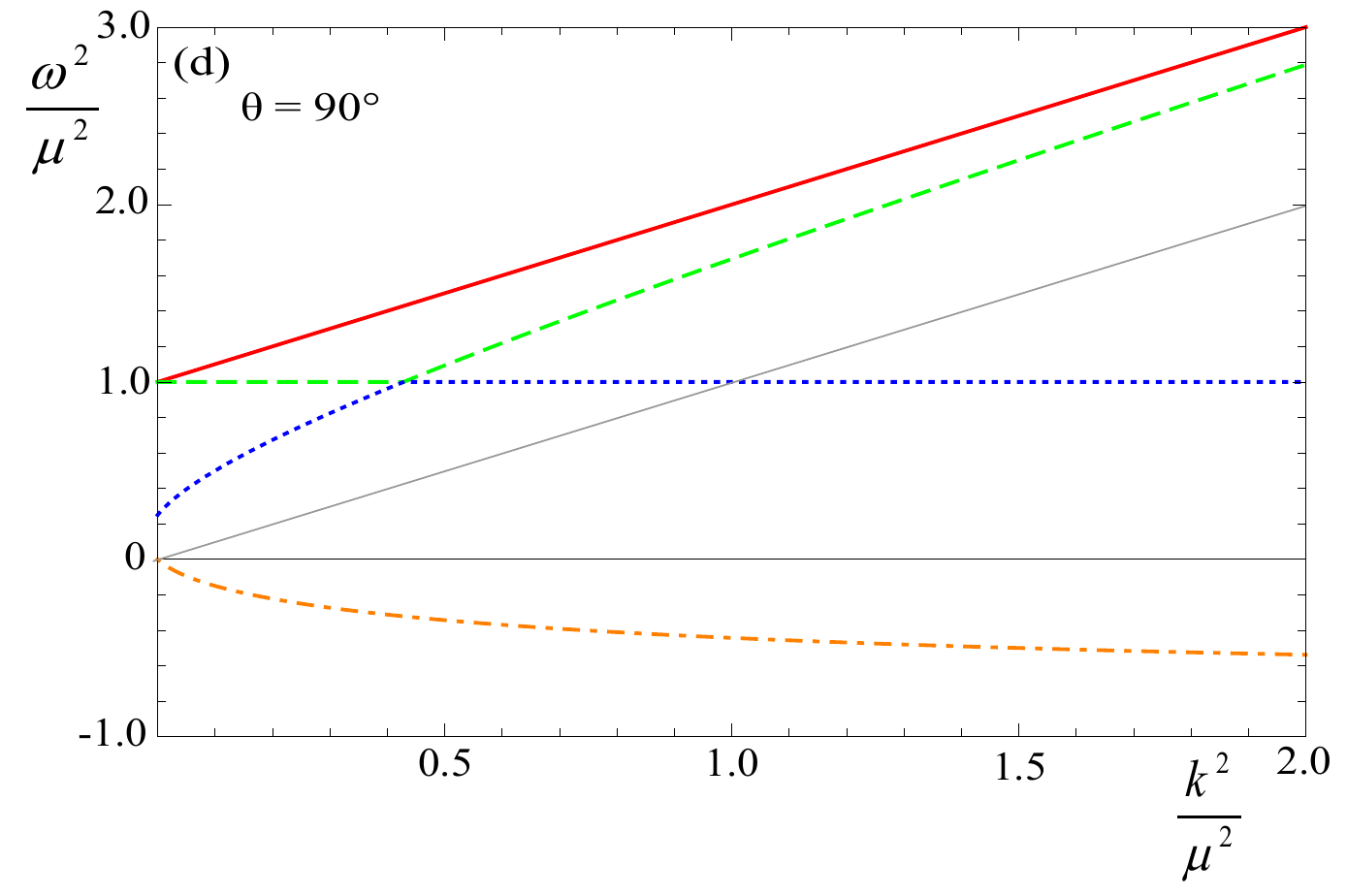}
\end{minipage}
\caption{(Color online) Dispersion curves $\omega^2({\bf k}) ~ vs. ~ {\bf k}^2$ at ${\bf u}^2 =3/4$ for four values of the angle $\theta$ equal $0^0$ (a), $30^0$ (b), $60^0$ (c), and $90^0$ (d). The red (solid) lines are for $\omega_a^2({\bf k})$, the green (dashed) for $\omega_1^2({\bf k})$, the blue (dotted) for $\omega_2^2({\bf k})$, and the orange (dashed-dotted) for $\omega_3^2({\bf k})$. The light cone is represented as a light gray line.
\label{fig-u2=3/4}}
\end{figure}

The real solutions of the cubic equation can be written down in the Vi\` ete's trigonometric form \cite{Bronshtein-Semendyayev-1985}
\be
\label{omega-n}
\omega_n^2({\bf k}) = 
2 \sqrt{\frac{-p}{3}} \cos\bigg[\frac{1}{3}\arccos\Big(\frac{3\sqrt{3} q }{2p^{\frac{3}{2}}}\Big) - \frac{2\pi n}{3}\bigg]
- \frac{a_2}{3 a_3} ,
\ee
where $n=1,2,3$ and 
\be
\label{p-q}
p \equiv \frac{3 a_3 a_1 - a_2^2}{3 a_3^2},
~~~~~~~~~~~~~~
q \equiv \frac{2 a_2^3 - 9 a_3 a_2 a_1 + 27 a_3^2 a_0}{27 a_3^3}  .
\ee
These formulas assume that $p<0$ and that the argument of the arccosine belongs to $[-1,1]$. These conditions are guaranteed as long as $\Delta = - a_3^4(4p^3 + 27 q^2) > 0$ which is the case under consideration. 

The complete set of dispersion curves of plasmons predicted by the formulas (\ref{A-mode}) and  (\ref{omega-n}) is shown in Fig.~\ref{fig-u2=3/4} for ${\bf u}^2=\frac{3}{4}$ and four different orientations of the wave vector ${\bf k}$.  As seen, the solution $\omega_3^2({\bf k})$ corresponds to the unstable and overdamped modes. We also observe in the figure that the green (dashed) line representing $\omega_1^2({\bf k})$ approaches the blue (dotted) line which refers to $\omega_2^2({\bf k})$. At $\theta = 90^0$ the lines hit each other but they do not cross. This is the phenomenon of {\it mode coupling} which is nicely explained in \S 64  of the textbook \cite{Landau-Lifshitz-1981}. 

How the general solutions (\ref{omega-n}) are related to those found in Sec.~\ref{sec-special-cases-k-u} for  ${\bf k} \perp  {\bf u}$ and ${\bf k} \parallel  {\bf u}$? In the former case, we have the relations:
\ba
\label{omega-11}
\omega_1^2 ({\bf k})  
=
\left\{ \begin{array}{ccc} 
\omega_0^2 (k)   \;\;\;\;\; & {\rm for} &  \;\;\;\;\; k <  \frac{\mu u}{\sqrt{1 + u^2}}  ,
\\ [2mm]
\omega_+^2 (k)   \;\;\;\;\; & {\rm for} &  \;\;\;\;\; k \ge \frac{\mu u}{\sqrt{1 + u^2}} ,
\end{array} \right.
\ea
\ba
\label{omega-22}
\omega_2^2 ({\bf k})  
=
\left\{ \begin{array}{ccc} 
\omega_+^2 (k)   \;\;\;\;\; & {\rm for} &  \;\;\;\;\;  k < \frac{\mu u}{\sqrt{1 + u^2}}  ,
\\ [2mm]
\omega_0^2 (k)   \;\;\;\;\; & {\rm for} &  \;\;\;\;\;  k \ge \frac{\mu u}{\sqrt{1 + u^2}} ,
\end{array} \right.
\ea
and $\omega_3^2 ({\bf k}) = \omega_-^2 (k)$. The crossing of the solutions $\omega_0^2 (k)$ and $\omega_+^2 (k)$ derived in Sec.~\ref{sec-k-perp-u} actually results from the limit $({\bf k} \cdot  {\bf u}) \rightarrow 0$.  The general solutions shown in Fig.~\ref{fig-u2=3/4} do not cross each other. 

The special solutions obtained in Sec.~\ref{sec-k-para-u} for  ${\bf k} \parallel  {\bf u}$ are related to the general solutions as:  $\omega_1^2 ({\bf k}) = \omega_a^2  ({\bf k})$,  $\omega_2^2 ({\bf k}) = \omega_+^2 (k)$ and  $\omega_3^2 ({\bf k}) = \omega_-^2 (k)$.

\section{Special cases ${\bf u}^2 =1$}
\label{sec-special-case-u2=1}

\begin{figure}[t]
\begin{minipage}{8.5cm}
\center
\includegraphics[width=\textwidth]{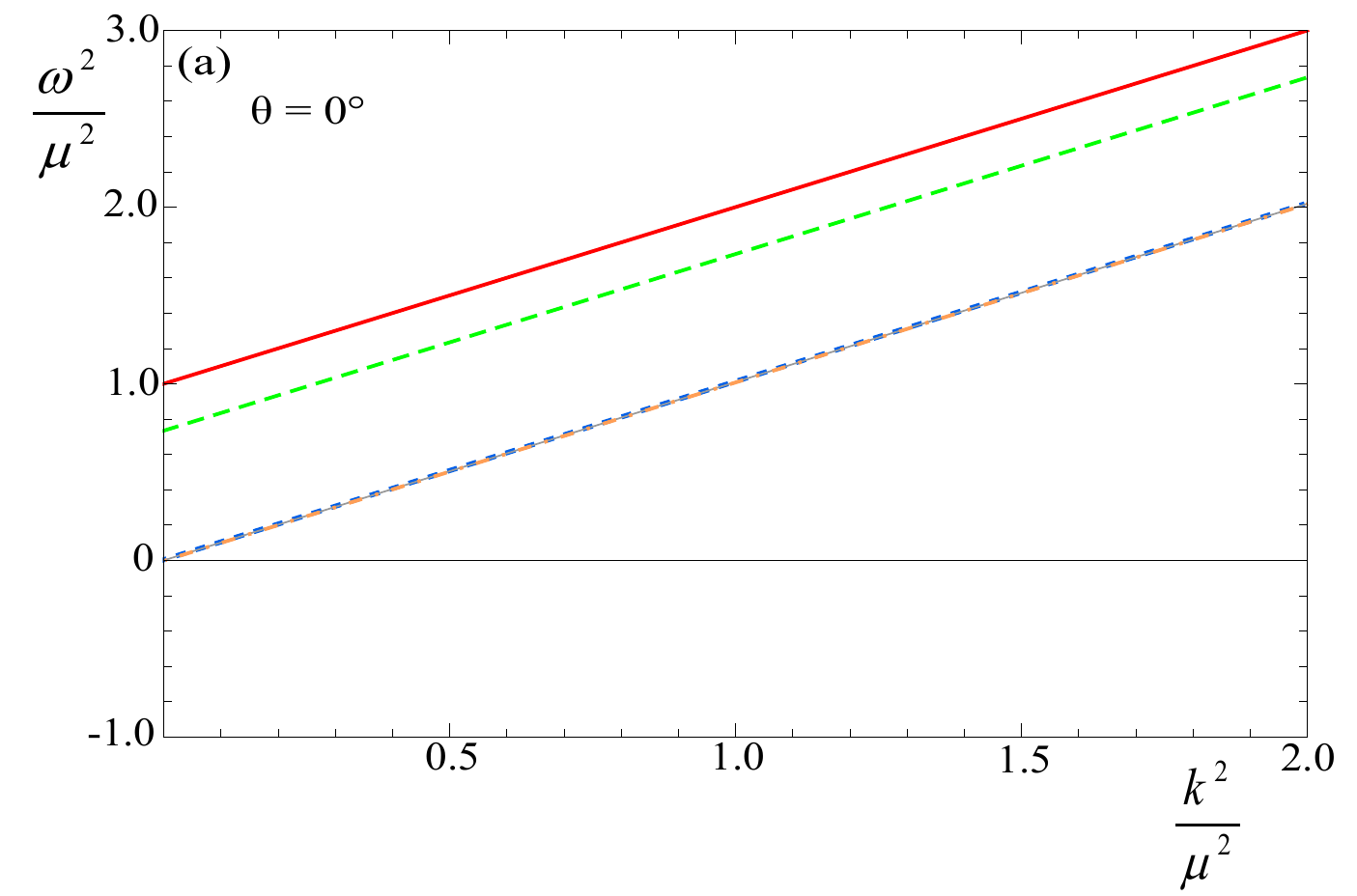}
\end{minipage}
\hspace{1mm}
\begin{minipage}{8.5cm}
\center
\includegraphics[width=\textwidth]{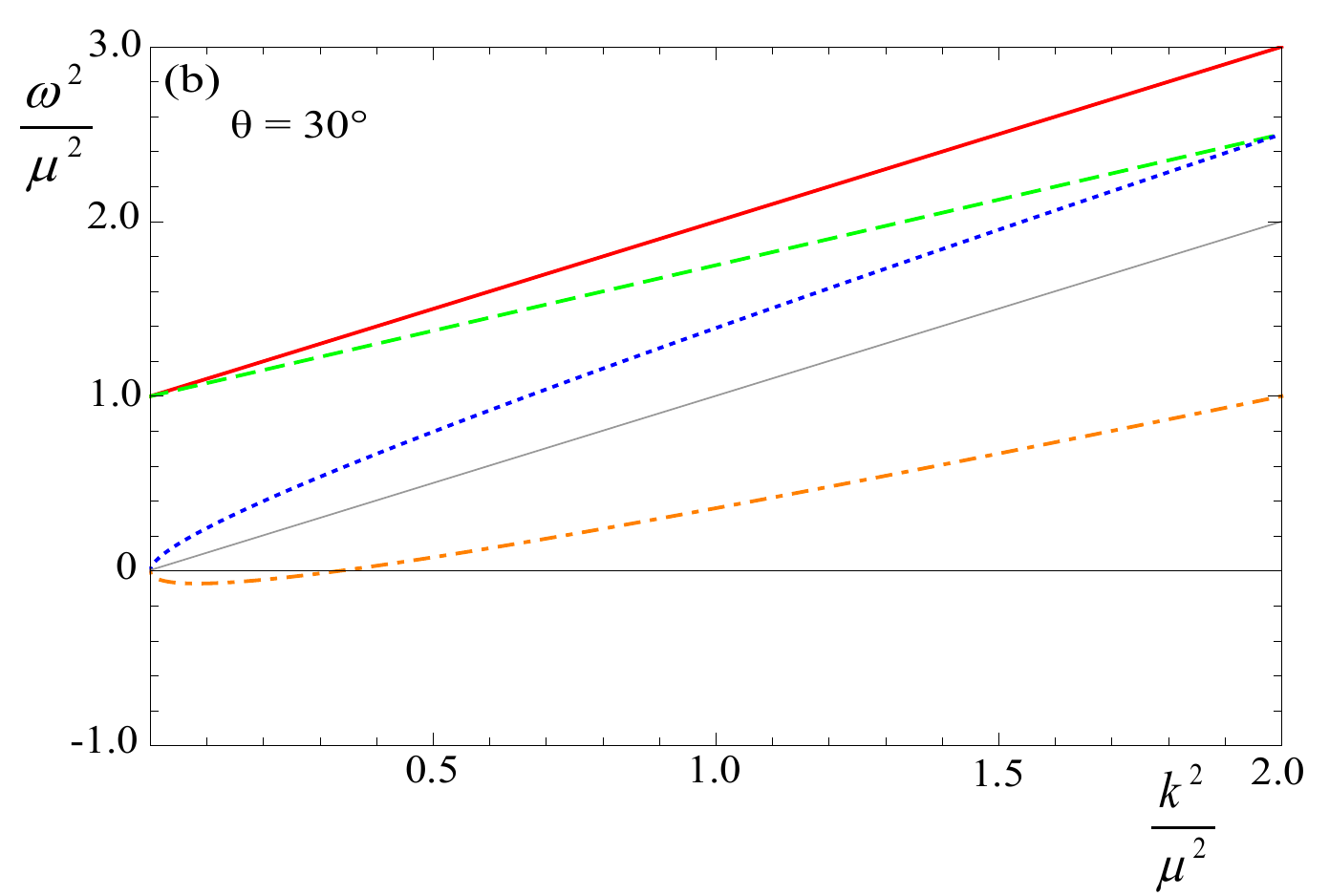}
\end{minipage}
\begin{minipage}{8.5cm}
\center
\includegraphics[width=\textwidth]{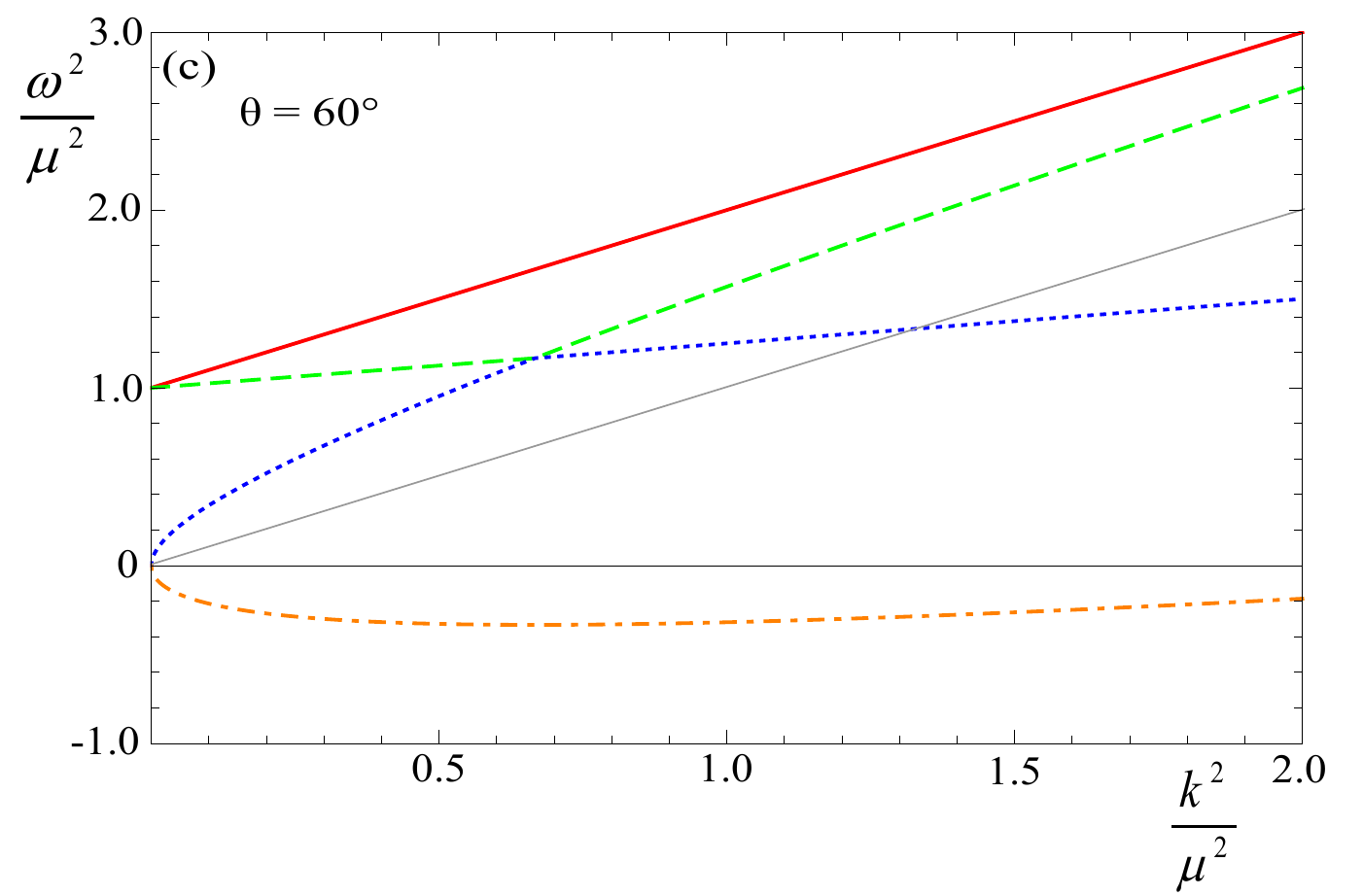}
\end{minipage}
\hspace{1mm}
\begin{minipage}{8.5cm}
\center
\includegraphics[width=\textwidth]{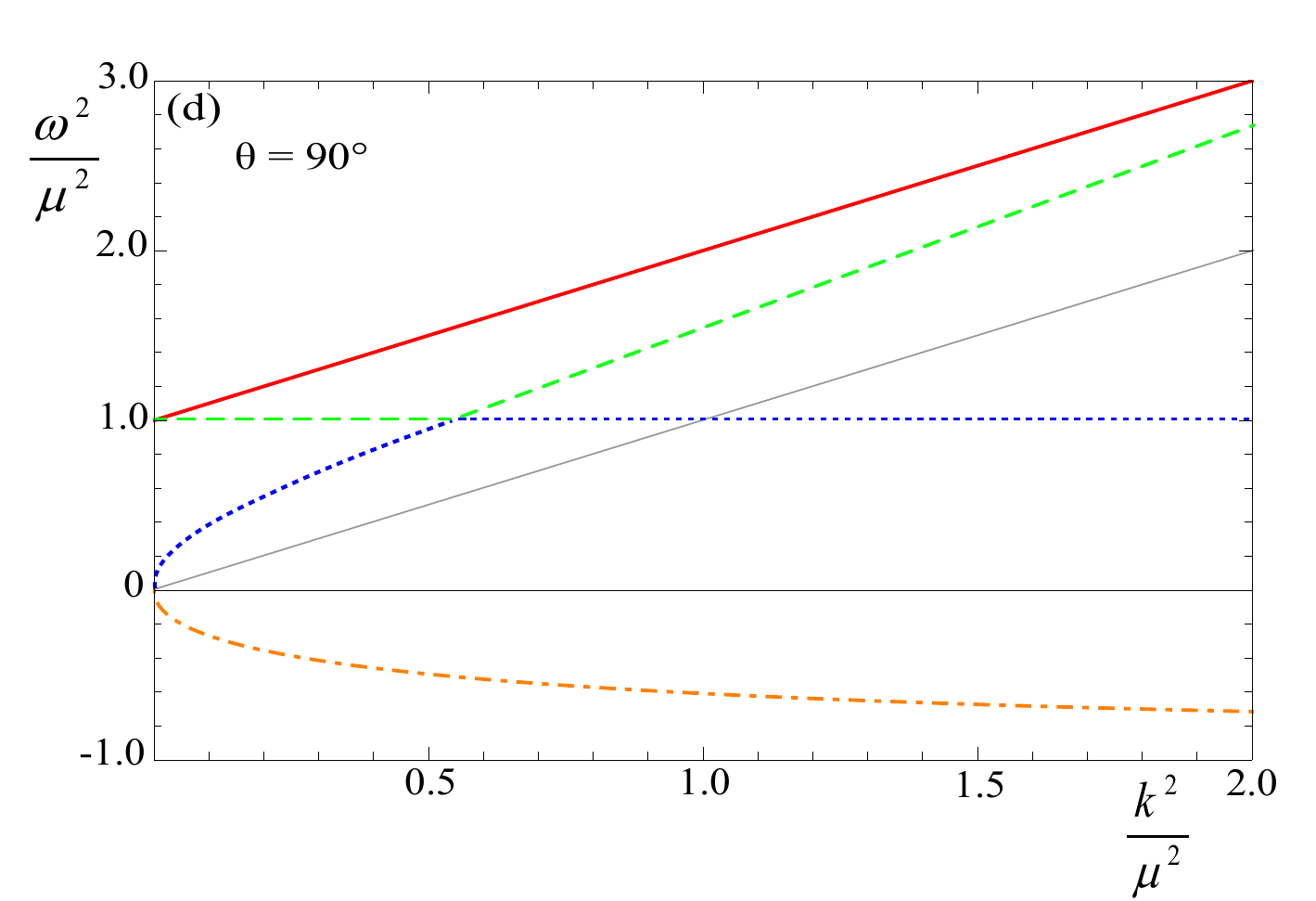}
\end{minipage}
\caption{(Color online) Dispersion curves $\omega^2({\bf k}) ~ vs. ~ {\bf k}^2$ at ${\bf u}^2 =1$ for four values of the angle $\theta$ equal $0^0$ (a), $30^0$ (b), $60^0$ (c), and $90^0$ (d). The red (solid) lines are for $\omega_a^2({\bf k})$, the green (dashed) for $\omega_1^2({\bf k})$, the blue (dotted) for $\omega_2^2({\bf k})$, and the orange (dashed-dotted) for $\omega_3^2({\bf k})$. The light cone is represented as a light gray line.
\label{fig-u2=1}}
\end{figure}

We consider here the case when the stream velocity $\bf {u}$ equals the speed of light. This case is interesting both from physical and mathematical points of view. The dielectric tensor of the two-stream system with ${\bf u}^2 = 1$ exactly coincides (under the replacement  ${\bf u}\rightarrow {\bf n}$ and $2\mu^2 \rightarrow m^2$) with that of the plasma with an extremely prolate (infinitely elongated in one direction) momentum distribution discussed in the study \cite{Carrington:2014bla}. Therefore, the plasmon spectra are obviously the same. Nevertheless, when the spectrum is found as a limit ${\bf u}^2 \rightarrow 1$, the mode crossing observed in the extremely prolate plasma gets a different physical meaning. As explained below, instead of mode crossing we rather have the extreme mode coupling mentioned in the previous section. 

The dispersion equation (\ref{cubic-eq}) ${\bf u}^2 = 1$ is also interesting mathematically. The form of the solutions (\ref{omega-n}) with the trigonometric and inverse trigonometric functions is required, if we deal with the so-called {\it casus irreducibilis}, when three real and distinct roots of a cubic equation cannot be expressed in terms of real radicals. However, a cubic equation, which has three real and distinct roots, can be sometimes reduced to a quadratic equation by means of the {\it rational root test}. Then, all three real roots of the cubic equation are expressed by real radicals and the Vi\` ete's trigonometric form is an unnecessary complication. 

Since the coefficient $a_3$ in Eq.~(\ref{cubic-eq}) equals unity, see Eq.~(\ref{a_3}), the rational root test suggests to look for a root of the equation among the factors of $a_0$ given by the formula (\ref{a_0}). When ${\bf u}^2 = 1$, there is a factor $\mu^2 + ({\bf k}\cdot {\bf u})^2$ which is indeed the root of the equation. Consequently, the cubic equation (\ref{cubic-eq}) is reduced to a  quadratic one, which is easily solved, and the three solutions read
\ba
\label{solutions-u1-1}
\omega_0^2({\bf k}) &=& \mu^2+({\bf k}\cdot {\bf u})^2,
\\[2mm]
\label{solutions-u1-pm}
\omega_\pm^2({\bf k}) &=&\frac{1}{2}\Big({\bf k}^2+({\bf k}\cdot {\bf u})^2\pm\sqrt{{\bf k}^4+({\bf k}\cdot {\bf u})^4+4\mu^2k^2-4\mu^2({\bf k}\cdot {\bf u})^2-2k^2({\bf k}\cdot {\bf u})^2}\Big).
\ea
The solutions $\omega_0^2({\bf k})$ and $\omega_+^2({\bf k})$ are positive for any ${\bf k}$ and consequently they represent  real modes. The modes $\omega_0({\bf k})$ and $\omega_+({\bf k})$ cross each other at $k = \frac{\mu}{\sqrt{2} \sin\theta}$. The solution $\omega_-^2({\bf k})$ is negative for $k <  \mu |\tan\theta|$ and positive otherwise. It represents the Weibel unstable mode and its  overdamped partner for sufficiently small wave vectors. When ${\bf k} \perp {\bf n}$ or $\theta = 90^\circ$, the unstable mode exists for all values of $k$. When ${\bf k} || {\bf n}$ or $\theta = 0^\circ$ the configuration is cylindrically symmetric and  there is no instability. This is the situation we have already encountered in Sec~\ref{sec-k-para-u} - the unstable longitudinal mode disappears as $u \rightarrow 1$.

Comparing the solutions (\ref{solutions-u1-1}) and (\ref{solutions-u1-pm}) to the general ones given by the formula (\ref{omega-n}), one realizes that
\ba
\label{omega-1}
\omega_1^2 ({\bf k})  
=
\left\{ \begin{array}{ccc} 
\omega_0^2 ({\bf k})   \;\;\;\;\; & {\rm for} &  \;\;\;\;\; k <  \frac{\mu}{\sqrt{2}\sin\theta},
\\ [2mm]
\omega_+^2 ({\bf k})   \;\;\;\;\; & {\rm for} &  \;\;\;\;\; k \ge \frac{\mu}{\sqrt{2}\sin\theta} ,
\end{array} \right.
\ea
\ba
\label{omega-2}
\omega_2^2 ({\bf k})  
=
\left\{ \begin{array}{ccc} 
\omega_+^2 ({\bf k})   \;\;\;\;\; & {\rm for} &  \;\;\;\;\;  k < \frac{\mu}{\sqrt{2}\sin\theta} ,
\\ [2mm]
\omega_0^2 ({\bf k})   \;\;\;\;\; & {\rm for} &  \;\;\;\;\;  k \ge \frac{\mu}{\sqrt{2}\sin\theta} ,
\end{array} \right.
\ea
and $\omega_3^2 ({\bf k}) = \omega_-^2 ({\bf k})$. Therefore, one sees that the crossing of the solutions $\omega_0({\bf k})$ and $\omega_+({\bf k})$ is actually an artifact of the limit ${\bf u}^2 \rightarrow 1$. The physical solutions are the combinations (\ref{omega-1}) and (\ref{omega-2}). 

The complete spectrum of plasmons, which includes $\omega_1^2 ({\bf k}), \; \omega_2^2 ({\bf k}), \; \omega_3^2 ({\bf k})$ and $\omega_a^2 ({\bf k})$, is shown in Fig.~\ref{fig-u2=1} for four different orientations of the wave vector ${\bf k}$. The qualitative difference, when compared to the case ${\bf u}^2 < 1$, occurs when $\theta \rightarrow 0$. Then, the solutions $\omega_2^2 ({\bf k})$ and $\omega_3^2 ({\bf k})$ merge into one double solution  $\omega_2^2 ({\bf k}) = \omega_3^2 ({\bf k})={\bf k}^2$.

\section{Summary and final remarks}
\label{sec-conclusions}

We have performed a systematic analysis of plasmons in the two-stream plasma system. The complete spectrum, which consists of four pairs of modes of opposite sign, has been found in a closed analytic form.  At any orientation of the wave vector and ${\bf u}^2 <1$, three pairs of modes are pure real and one pair is pure imaginary for sufficiently small wave vectors. When the wave vector is bigger than the critical one, all four pairs of modes are real. Among the imaginary modes, one is unstable and one is overdamped. An interesting feature of the unstable mode is that when the orientation of the wave vector changes from parallel to the stream velocity to perpendicular, the mode smoothly changes from longitudinal to transverse. In the first case, we have the electrostatic two-stream instability and in the latter one the filamentation or Weibel instability. When ${\bf u}^2 \rightarrow 1$, the electrostatic unstable mode disappears.  Therefore, when ${\bf u}^2 =1$ and ${\bf k} \parallel {\bf u}$, the whole spectrum of modes is real. 

In the limiting cases ${\bf k} \perp {\bf u}$ and ${\bf u}^2 = 1$, one seems to observe mode crossing. However, a confrontation of the special case solutions with the general ones shows that there is no mode crossing but rather extreme mode coupling when the dispersion curves hit each other but they do not cross. 

A mathematical structure of the plasmon spectrum is also interesting. In the general case we deal with the {\it casus irreducibilis} when the three solutions of the cubic dispersion equation cannot be expressed in terms of real radicals. Therefore, the solutions are given by the trigonometric and inverse trigonometric functions. However, in the extreme cases ${\bf k} \perp {\bf u}$, ${\bf k} \parallel {\bf u}$ and ${\bf u}^2 = 1$, the cubic equation is reduced to the quadratic one and the solutions are expressed in terms of real quadratic radicals. 

The two-stream configuration repeatedly occurs in the physics of electromagnetic plasma and thus our results are hopefully of some practical interest.  In case of the quark-gluon plasma produced in relativistic heavy-ion collisions, the two-stream system is rather irrelevant. However, such a system can be treated as a simple but nontrivial model of an unstable plasma. We have actually used the model to study the energy-loss of a high-energy parton \cite{Carrington:2012hv} and it can be also applied to other problems. 

\section*{Acknowledgments}

We are very grateful to Margaret Carrington for discussions and critical reading of the manuscript. This work was partially supported by the Polish National Science Centre under grant 2011/03/B/ST2/00110. 


\end{document}